\documentclass[conference]{IEEEtran}
\usepackage[T1]{fontenc}
\usepackage[utf8]{inputenc}
\usepackage{amssymb}
\usepackage[pdftex]{graphicx}
\usepackage{amsmath}
\usepackage{amsthm}
\usepackage{mathtools}
\usepackage[noadjust]{cite}
\usepackage[acronym]{glossaries}
\usepackage{siunitx}
\usepackage{tikz}
\usepackage{flushend}
\usepackage{multirow}
\usepackage{pgfplots}
\usepackage{nicefrac}
\usepackage{fancyref}
\usepackage[capitalise]{cleveref}
\pgfplotsset{compat=1.17}
\usetikzlibrary{arrows,chains,matrix,positioning,scopes}
\usetikzlibrary{calc}
\usetikzlibrary{fit,shapes}
\usepackage{balance}
\newacronym{NPRACH}{NPRACH}{narrowband physical random-access channel}
\newacronym{ToA}{ToA}{time of arrival}
\newacronym{CFO}{CFO}{carrier frequency offset}
\newacronym{NBIoT}{NB-IoT}{narrowband internet of things}
\newacronym{5GNR}{5G NR}{5G New Radio}
\newacronym{3GPP}{3GPP}{3rd Generation Partnership Project}
\newacronym{UMi}{UMi}{urban microcell}
\newacronym{RMSE}{RMSE}{root-mean-square error}
\newacronym{NN}{NN}{neural network}
\newacronym{BS}{BS}{base station}
\newacronym{UE}{UE}{user equipment}
\newacronym{SG}{SG}{symbol group}
\newacronym{CP}{CP}{cyclic prefix}
\newacronym{OFDM}{OFDM}{orthogonal frequency division multiplexing}
\newacronym{FFT}{FFT}{fast Fourier transform}
\newacronym{AWGN}{AWGN}{additive white Gaussian noise}
\newacronym{DFT}{DFT}{discrete Fourier transform}
\newacronym{FNR}{FNR}{false negative rate}
\newacronym{FPR}{FPR}{false positive rate}
\newacronym{RG}{RG}{resource grid}
\newacronym{RE}{RE}{resource element}
\newacronym{SNR}{SNR}{signal-to-noise ratio}
\newacronym{1D}{1D}{one-dimensional}
\newacronym{MLP}{MLP}{multilayer perceptron}
\newacronym{BCE}{BCE}{binary cross-entropy}
\newacronym{KL}{KL}{Kullback–Leibler}
\newacronym{SGD}{SGD}{stochastic gradient descent}
\newacronym{ppm}{ppm}{parts-per-million}
\newacronym{ICI}{ICI}{inter-carrier interference}
\newacronym{GNN}{GNN}{graph neural network}
\newacronym{BP}{BP}{belief propagation}
\newacronym{FEC}{FEC}{forward error correction}
\newacronym{LDPC}{LDPC}{low-density parity-check}
\newacronym{HDPC}{HDPC}{high-density parity-check}
\newacronym{SCL}{SCL}{successive cancellation list}
\newacronym{SC}{SC}{successive cancellation}
\newacronym{URLLC}{URLLC}{ultra-reliable low-latency communications}
\newacronym{APP}{APP}{a posterior probability}
\newacronym{MIMO}{MIMO}{multiple-input multiple-output}
\newacronym{CNN}{CNN}{convolutional neural network}
\newacronym{BER}{BER}{bit error rate}
\newacronym{BPSK}{BPSK}{binary phase shift keying}
\newacronym{LLR}{LLR}{log-likelihood ratio}
\newacronym{FN}{FN}{factor node}
\newacronym{VN}{VN}{variable node}
\newacronym{CN}{CN}{check node}
\newacronym{MPNN}{MPNN}{message passing neural network}

\newacronym{AI}{AI}{artificial intelligence}
\newacronym{ML}{ML}{machine learning}
\newacronym{SISO}{SISO}{single input single output}
\newacronym{PRB}{PRB}{physical resource block}
\newacronym{PUSCH}{PUSCH}{physical uplink shared channel}

\newacronym{MUMIMO}{MU-MIMO}{multi-user multiple-input multiple-output}
\newacronym{BICM}{BICM}{bit-interleaved coded modulation}
\newacronym{QAM}{QAM}{quadrature amplitude modulation}
\newacronym{LMMSE}{LMMSE}{linear minimum mean square error}
\newacronym{CSI}{CSI}{channel state information}
\newacronym{SIMO}{SIMO}{single-input multiple-output}
\newacronym{CGNN}{CGNN}{convolutional and graph neural network}
\newacronym{BLER}{BLER}{block error rate}
\newacronym{LS}{LS}{least squares}
\newacronym{PE}{PE}{positional encoding}
\newacronym{relu}{ReLU}{rectified linear unit}
\newacronym{RB}{RB}{resource block}
\newacronym{CGGNN}{CGGNN}{convolutional graph neural network}
\newacronym{DMRS}{DMRS}{demodulation reference signal}
\newacronym{IoT}{IoT}{internet of things}
\newacronym{ADAM}{ADAM}{adaptive momentum}
\newacronym{TBLER}{TBLER}{transport block error rate}
\newacronym{MCS}{MCS}{modulation and coding scheme}
\newacronym{TDL}{TDL}{tapped delay line}
\newacronym{CDM}{CDM}{code division multiplexing}
\newacronym{FLOP}{FLOP}{floating point operation}
\newacronym{PHY}{PHY}{physical layer}

\newacronym{ULA}{ULA}{uniform linear array}
\newacronym{NRX}{NRX}{neural receiver}
\newacronym{Var-MCS-NRX}{Var-MCS NRX}{variable-MCS NRX}
\newacronym{UL}{UL}{uplink}
\newacronym{DL}{DL}{downlink}
\newacronym{MSE}{MSE}{mean squared error}
\newacronym{CIR}{CIR}{channel impulse response}
\newacronym{iid}{iid}{independent and identically distributed}

\newacronym{AI-RAN}{AI-RAN}{AI radio access network}
\newacronym{ISAC}{ISAC}{integrated sensing and communications}
\newacronym{RIS}{RIS}{reflecting intelligent surfacces}

\definecolor{mittelblau}{RGB}{0, 126, 198}
\definecolor{violettblau}{cmyk}{0.9, 0.6, 0, 0}
\definecolor{rot}{RGB}{238, 28 35}
\definecolor{apfelgruen}{RGB}{140, 198, 62}
\definecolor{gelb}{RGB}{1, 221, 0}
\definecolor{orange}{RGB}{244, 111, 33}
\definecolor{pink}{RGB}{237, 0, 140}
\definecolor{lila}{RGB}{128, 10, 145}
\definecolor{hellgrau}{RGB}{224, 224, 224}
\definecolor{mittelgrau}{RGB}{128, 128, 128}
\definecolor{dunkelgrau}{RGB}{80,80,80}
\definecolor{anthrazit}{RGB}{19, 31, 31}

\pgfplotscreateplotcyclelist{corporate_colours_markers}{%
rot, every mark/.append style={fill=.!80!rot},mark=*\\%
mittelblau, every mark/.append style={fill=.!80!mittelblau},mark=square*\\%
apfelgruen, every mark/.append style={fill=.!80!apfelgruen},mark=triangle*\\%
violettblau, every mark/.append style={fill=.!80!violettblau},mark=otimes*, densely dotted\\%
orange, mark=star, densely dotted\\%
pink, every mark/.append style={fill=.!80!pink},mark=diamond*, densely dashed\\%
lila, mark=|, densely dashed\\%
gelb, every mark/.append style={fill=.!80!gelb},mark=pentagon*\\%
hellgrau, mark=text,text mark=p\\%
anthrazit, mark=text,text mark=a\\%
}

\pgfplotscreateplotcyclelist{corporate_colours_no_markers}{%
rot, every mark/.append style={fill=.!80!rot},mark=none\\%
mittelblau, every mark/.append style={fill=.!80!mittelblau},mark=none\\%
apfelgruen, every mark/.append style={fill=.!80!apfelgruen},mark=none\\%
violettblau, every mark/.append style={fill=.!80!violettblau},mark=none, densely dotted\\%
orange, mark=none, densely dotted\\%
pink, every mark/.append style={fill=.!80!pink},mark=none, densely dashed\\%
lila, mark=none, densely dashed\\%
gelb, every mark/.append style={fill=.!80!gelb},mark=none\\%
hellgrau, mark=text,text mark=none\\%
anthrazit, mark=text,text mark=none\\%
}

\pgfplotscreateplotcyclelist{corporate_colours_no_markers1}{%
anthrazit, mark=none\\%
rot, every mark/.append style={fill=.!80!rot},mark=none\\%
violettblau, every mark/.append style={fill=.!80!violettblau},mark=none\\%
apfelgruen, every mark/.append style={fill=.!80!apfelgruen},mark=none\\%
orange, mark=none\\%
pink, every mark/.append style={fill=.!80!pink},mark=none\\%
mittelblau, every mark/.append style={fill=.!80!mittelblau},mark=none\\%
lila, mark=none, densely dashed\\%
gelb, every mark/.append style={fill=.!80!gelb},mark=none\\%
gray, mark=text,text mark=none\\%
}

\pgfplotscreateplotcyclelist{corporate colours markers double}{%
rot, every mark/.append style={fill=.!80!rot},mark=*\\%
rot, dashed, every mark/.append style={fill=.!80!rot,solid},mark=*\\%
mittelblau, every mark/.append style={fill=.!80!mittelblau},mark=square*\\%
mittelblau, dashed, every mark/.append style={fill=.!80!mittelblau,solid},mark=square*\\%
apfelgruen, every mark/.append style={fill=.!80!apfelgruen},mark=triangle*\\%
apfelgruen, dashed, every mark/.append style={fill=.!80!apfelgruen,solid},mark=triangle*\\%
orange, mark=star\\%
orange, dashed, mark=star\\%
pink, every mark/.append style={fill=.!80!pink},mark=diamond*\\%
pink, dashed, every mark/.append style={fill=.!80!pink,solid},mark=diamond*\\%
violettblau, every mark/.append style={fill=.!80!violettblau},mark=otimes*\\%
violettblau, dashed, every mark/.append style={fill=.!80!violettblau,solid},mark=otimes*\\%
lila, mark=|\\%
lila, dashed, mark=|\\%
gelb, every mark/.append style={fill=.!80!gelb},mark=pentagon*\\%
gelb, dashed, every mark/.append style={fill=.!80!gelb,solid},mark=pentagon*\\%
}

\IEEEoverridecommandlockouts 

\begin{document}

\title{Sionna Research Kit: A GPU-Accelerated Research Platform for AI-RAN}

\author{
\IEEEauthorblockN{Sebastian Cammerer, Guillermo Marcus, Tobias Zirr, Fay\c{c}al A\"{i}t Aoudia, \\Lorenzo Maggi, Jakob Hoydis, and Alexander Keller}
\IEEEauthorblockA{NVIDIA, contact: scammerer@nvidia.com}

\thanks{This work has received financial support from the European Union under Grant Agreement 101096379 (CENTRIC).  Views and opinions expressed are however those of the author(s) only and do not necessarily reflect those of the European Union or the European Commission (granting authority).  Neither the European Union nor the granting authority can be held responsible for~them.}
}

\maketitle

\glsresetall

\begin{abstract}
We introduce the NVIDIA Sionna Research Kit,  a GPU-accelerated research platform for developing and testing AI/ML algorithms in 5G NR cellular networks. Powered by the NVIDIA Jetson AGX Orin, the platform leverages accelerated computing to deliver high throughput and real-time signal processing, while offering the flexibility of a software-defined stack. Built on OpenAirInterface (OAI) \cite{nikaein2014openairinterface},  it unlocks a broad range of research opportunities. These include developing 5G NR and O-RAN compliant algorithms, collecting real-world data for AI/ML training, and rapidly deploying innovative solutions in a very affordable testbed. Additionally, AI/ML hardware acceleration promotes the exploration of use cases in edge computing and AI radio access networks (AI-RAN).
To demonstrate the capabilities, we deploy a real-time neural receiver—trained with NVIDIA Sionna and using the NVIDIA TensorRT library for inference—in a 5G NR cellular network using commercial user equipment. The code examples will be made publicly available, enabling researchers to adopt and extend the platform for their own projects.
\end{abstract}

\glsresetall
\section{Background}

\Gls{AI} for wireless communications has received significant attention from  both academia \cite{o2017introduction} and industry \cite{lin2023overview}. Besides the potential to deliver superior reliability and accuracy as compared to many traditional physical layer algorithms, \gls{AI} facilitates the development of novel concepts such as site-specific adaptation \cite{wiesmayr2024nrx}, end-to-end learning \cite{o2017introduction}, and semantic communications \cite{qin2021semantic}. Moreover, the AI-RAN Alliance\footnote{{https://ai-ran.org/}} envisions that generative AI applications will drive future 6G system requirements, such as edge AI offloading and low-latency communications.

Though many concepts and algorithms have been proposed in the literature, so far little has been demonstrated in hardware testbeds and under real-time conditions. Certainly, the stringent latency and throughput requirements of wireless systems impose strict constraints on \gls{NN} architectures, effectively limiting their size and depth. Thus, deploying and validating AI components in the physical layer of a real cellular system under realistic latency conditions remains an open and yet exciting research challenge.

Hence, the development and evaluation of novel AI-RAN algorithms requires: (a)~affordable hardware platforms, (b)~programmable hardware accelerators for real-time signal processing as well as \gls{AI} offloading, and (c) a low  barrier-to-entry considering the system complexity of modern wireless communication standards. Often, these requirements have kept academic researchers from engaging in practical system-level experimentation. The software-defined open stack of the Sionna Research Kit bridges this gap, enabling researchers to test their algorithms in an operational 5G NR system, boosting result credibility and unlocking new research opportunities.

\begin{figure}[t]
    \centering
    \includegraphics[width=0.92\columnwidth]{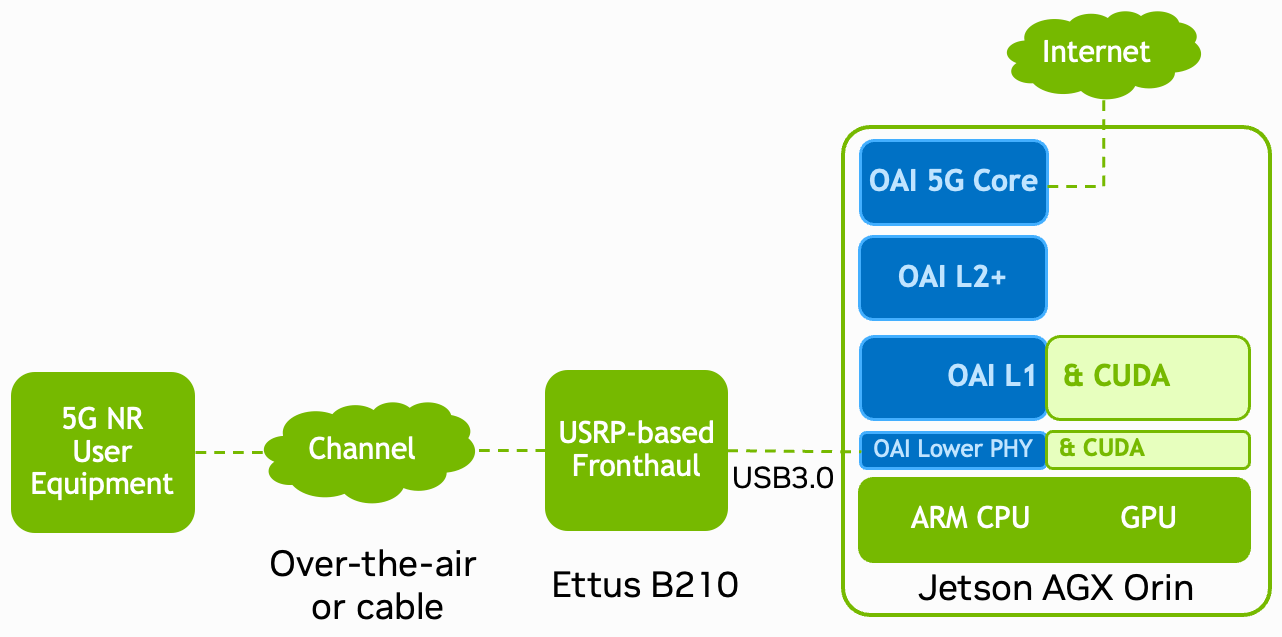}
    \caption{Schematic of the demo setup, consisting of an NVIDIA Jetson AGX Orin, an USRP B210 by Ettus Research, and a commercial Quectel RM520N-GL 5G NR modem.}
    \label{fig:system_overview}
\end{figure}

\section{Sionna Research Kit}

The Sionna Research Kit\footnote{{https://nvlabs.github.io/sionna/rk/index.html}} relies on the NVIDIA Jetson AGX Orin platform.\footnote{{https://www.nvidia.com/en-us/autonomous-machines/embedded-systems/}} GPU acceleration using NVIDIA CUDA\footnote{{https://developer.nvidia.com/cuda-toolkit}} enables real-time signal processing with high throughput while preserving the flexibility of a software-defined stack.

In principle, approaches to hardware acceleration of physical layer signal processing may be classified as either (a) look-aside acceleration or (b) inline acceleration \cite{kundu2023hardware}.
Look-aside acceleration offloads specific tasks from the CPU to an accelerator—such as an ASIC, FPGA, or GPU—which then processes them asynchronously.
While look-aside acceleration may include fixed function hardware for specific function blocks (e.g., for \gls{LDPC} decoding), moving data to and from the accelerator may add latency.
In contrast, inline acceleration integrates directly into the data processing pipeline. 
An advantage of the NVIDIA Jetson platform is that both CPU and GPU share the same unified memory, thus eliminating the data copying overhead.

\subsection{Case Study: Neural Receiver Under Real-Time Constraints}

We have developed a prototype of a 5G NR standard-compliant neural receiver \cite{wiesmayr2024nrx} that replaces parts of the physical layer signal processing with machine-learned components. The architecture has been carefully optimized to ensure real-time inference capabilities. For details, see \cite{cammerer2023neural} and \cite{wiesmayr2024nrx}.

\begin{figure}[t]
    \centering
    \includegraphics[width=0.9\columnwidth]{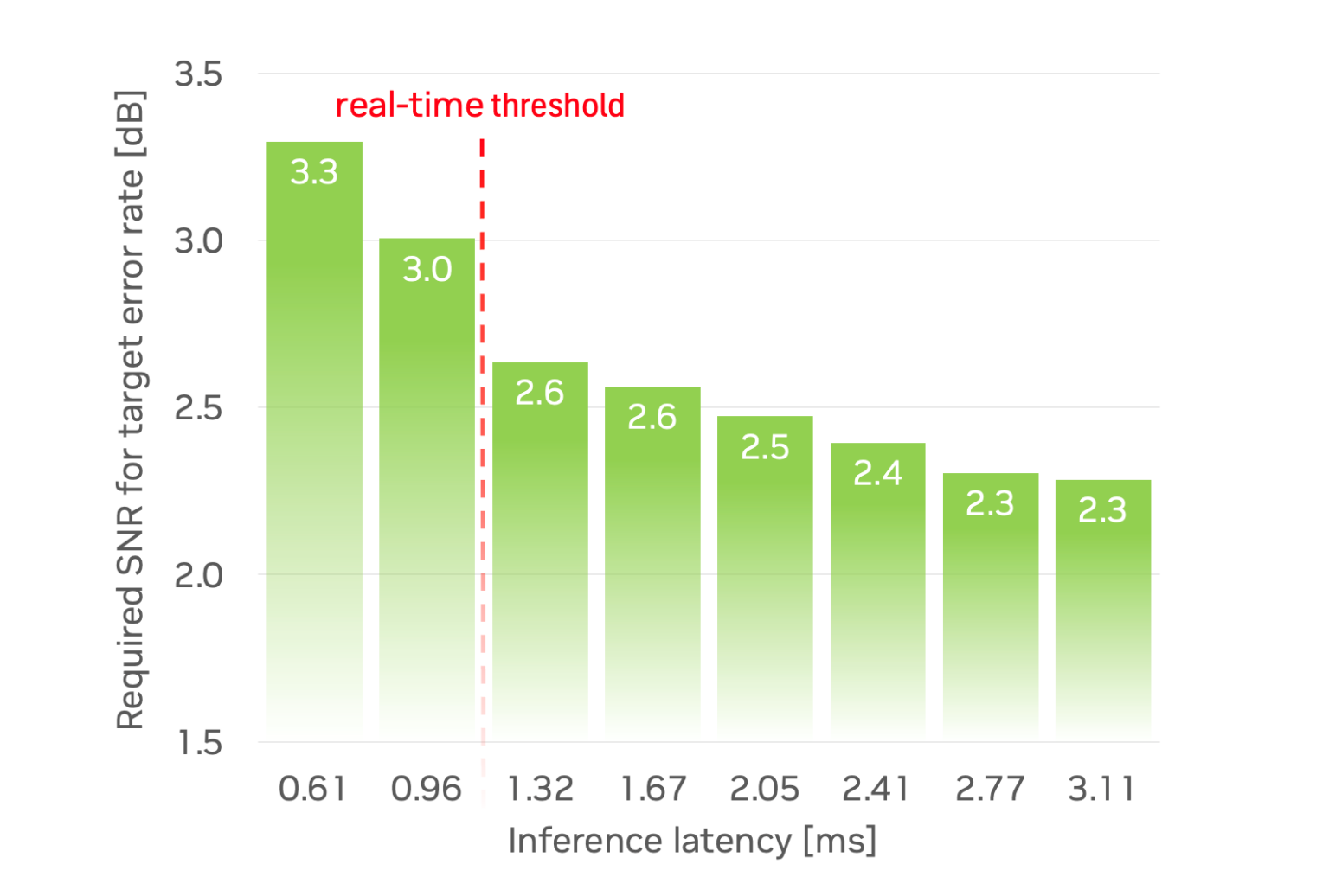}
    \vspace*{-0.2cm}
    \caption{Performance evaluation of the NRX \cite{wiesmayr2024nrx}, varying its network depth and, hence, the inference latency.  Figure taken from {https://developer.nvidia.com/blog/real-time-neural-receivers-drive-ai-ran-innovation/}.}
    \label{fig:nrx_perf}
\end{figure}

As an example, we showcase the deployment the neural receiver in a 5G cellular network. The receiver is trained with NVIDIA Sionna~\cite{hoydis2022sionna} and implemented using the NVIDIA TensorRT inference library.
Fig.~\ref{fig:nrx_perf} illustrates that the required SNR to achieve a target \gls{BLER} of 0.1 depends on the network depth and hence the inference latency. These results demonstrate that real-time considerations significantly influence the performance and, thus, the design of AI/ML components in wireless systems.

\subsection{Case Study: CUDA-accelerated LDPC Decoding}

In a second example, we demonstrate GPU offloading in wireless systems by implementing a CUDA-accelerated LDPC decoder which seamlessly integrates into the OAI stack. To avoid the latency overhead of data transfer between the CPU and the GPU, we carefully optimize the caching behavior of the implemented CUDA kernels. While some data movement between CPU and GPU still occurs, the unified memory architecture of the Jetson AGX Orin significantly reduces the overhead of copying data.

\section{Demo Setup \& Hardware Requirements}

\begin{figure}[t]
    \centering
    \includegraphics[width=\linewidth]{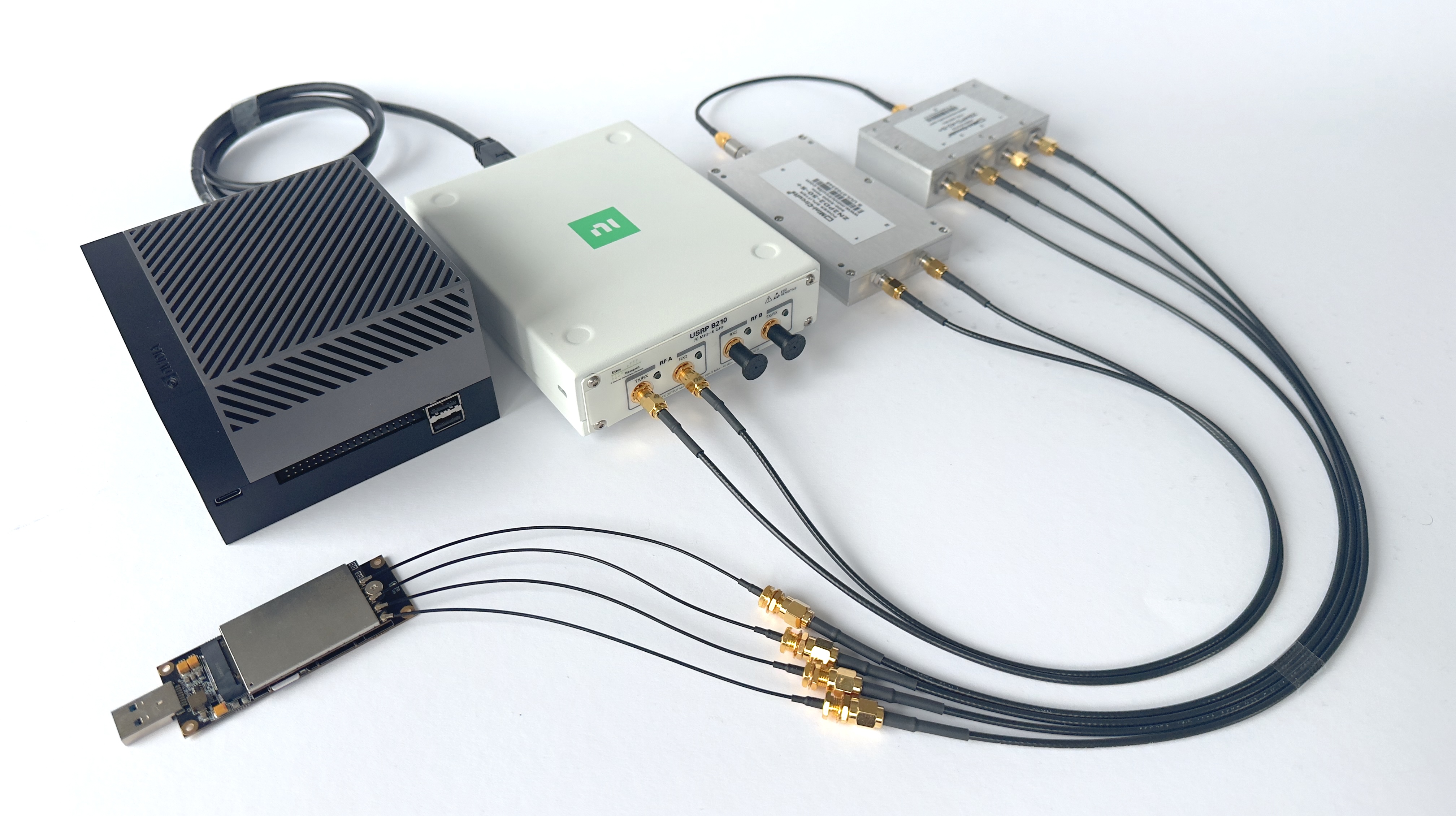}
    \caption{Photograph of the hardware components of the demo setup.}
    \label{fig:setup}
\end{figure}

Fig.~\ref{fig:system_overview} shows the schematic of the demo and Fig.~\ref{fig:setup} reveals the hardware components consisting of:
\begin{itemize}
    \item \textbf{NVIDIA Jetson AGX Orin:} A GPU-accelerated embedded platform enabling real-time signal processing and AI/ML inference.
    \item \textbf{Ettus Research USRP B210 SDR:} A flexible SDR front-end interfacing with the software-defined 5G NR stack.
    \item \textbf{Quectel RM520N-GL 5G Modem:} A commercial 5G NR modem serving as the \gls{UE}, highlighting the demo's standard compliance.
\end{itemize}

Both experiments will be made publicly available and can serve as blueprints for implementing novel research prototypes. Additionally, tutorials on real-world data acquisition and a TensorRT accelerated neural demapper are provided.  This enables researchers to acquire the necessary data for training novel AI-RAN algorithms and deploying their trained models using an inline AI/ML accelerated computing platform.

\section{Conclusion}

The Sionna Research Kit enables researchers to develop, deploy, test, and validate novel AI/ML algorithms in a 5G NR cellular network, including software-defined user equipment (UE). Its unique combination of the OpenAirInterface software-defined stack and the NVIDIA Jetson AGX Orin platform allows for real-time signal processing and AI/ML inference.
As such, we believe the Sionna Research Kit is a step towards practical prototyping of the next-generation AI-RAN.

\bibliographystyle{IEEEtran}
\bibliography{IEEEabrv, bibliography/VIPabbrv, bibliography/bibliography}

\end{document}